\documentclass[twocolumn,pra,showpacs,superscriptaddress]{revtex4}
\usepackage{graphicx}
\usepackage{dcolumn}
\usepackage{bm}
\usepackage{leftidx}

\begin{document}
\title{Complex quantum network model of energy transfer in photosynthetic complexes}

\author{Bao-quan  Ai}
\affiliation{Laboratory of Quantum Information Technology and
SPTE,South China Normal University, Guangzhou, China}
\author{Shi-Liang Zhu}\email{shilzhu@yahoo.com.cn}
\affiliation{Laboratory of Quantum Information Technology and
SPTE,South China Normal University, Guangzhou, China}

\begin{abstract}
The quantum network model with real variables is usually used to
describe the excitation energy transfer (EET) in the
Fenna-Matthews-Olson(FMO) complexes. In this paper we add the
quantum phase factors to the hopping terms and find that the
quantum phase factors play an important role in the EET. The
quantum phase factors allow us to consider the space structure of
the pigments. It is found that phase coherence within the
complexes would allow quantum interference to affect the dynamics
of the EET. There exist some optimal phase regions where the
transfer efficiency takes its maxima, which indicates that when
the pigments are optimally spaced, the exciton can pass through
the FMO with perfect efficiency. Moreover, the optimal phase
regions almost do not change with the environments. In addition,
we find that the phase factors are useful in the EET just in the
case of multiple-pathway. Therefore, we demonstrate that, the
quantum phases may bring the other two factors, the optimal space
of the pigments and multiple-pathway, together to contribute the
EET in photosynthetic complexes with perfect efficiency.
\end{abstract}

 \maketitle

\section{introduction}

Photosynthesis provides chemical energy for almost all life on
Earth. The initial step of photosynthesis involves absorption of
light by the so-called light-harvesting antennae complexes, and
funneling of the resulting electronic excitation to the
photosynthetic reaction center.  Recent work has reported that
quantum theory governs the exciton transfer in some
light-harvesting complexes that harness the absorbed energy with
almost $100\%$ efficiency \cite{a0,a1,a2,a3,a4,a5,a6}.  The
experimental evidence \cite{a1,a2,a3,a4,a6} showing long-lived
quantum coherences in this energy transport in several
photosynthetic light harvesting complexes suggests that coherence
may play an important role in the function of these systems. These
observations have generated considerable interest in understanding
the possibly functional role of quantum coherence effects in the
remarkably efficient excitation energy
transfer in photosynthetic complexes. \\

 The experimental achievements have motivated a number of
theoretical works
\cite{a7,b1,b2,b3,b4,b5,b6,b7,b8,b9,b10,b11,yi,b12,b13,b14,baa,b15,b16,Ringsmuth,Jing,Scholak,Cao}
that consider the photosynthetic complex as a quantum system, and
try to analyze the basic mechanisms that explain the phenomena
observed in the experiments. A full quantum dynamic framework
becomes necessary for studying coherent energy transfer.  Typical
quantum theories are, the quantum network model
\cite{b1,b2,b3,b4,b6,b7,b8,b9,b10,b11,yi,Ringsmuth,Scholak}, the
hierarchic equation \cite{a7,b5,b12,b13}, the generalized
Bloch-Redfield \cite{b14} equation, the renormalization
group methods \cite{baa}, and the mixed quantum-classical
method\cite{Cao}. Some recent theories \cite{Christensson}
 can even successfully predict the long-lived quantum coherent
phenomenon. However, all these models cannot describe the space
distribution of the pigments. As we know, the space distribution
of the pigments is very important for exciton energy transfer and
the experimental evidence also shows that the optimized space
distribution of the pigments is one factor for prefect energy
transfer in light-harvesting complexes \cite{a1,a2}. Therefore, it
is necessary to set up a quantum model in which the space
distribution of the pigments is considered.

In this paper, we set up a quantum network model by adding the
quantum phase factors to the two-body interactions to describe the
exciton (the energy carrier) transfer in the FMO complex. The
quantum phases are determined by the spatial structure of the
pigments in photosynthetic complexes, such as the length of the
pigment, the barriers and the distance between pigments. The
quantum network in the absence of the phase factors has been used
to study the EET in the photosynthetic
complexes\cite{b1,b2,b3,b4,b6,b7,b8,b9,b10,b11,yi}. Some
interesting results, such as noises may enhance the EET, and the
EET in a quantum model may be larger than that of a classical
model, are obtained. Compared with those studies, we find that the
newly added quantum phase factors play a key role in the EET  and
there exist optimized phases at which the transfer efficiency is
maximal. Furthermore, we find that the phase factors affect the
EET just in the form of the phase difference in a closed loop.
Although there may be many phase factors in the coupling terms of
the system Hamiltonian, only $N_p-1$ are independent variables
where $N_p$ is the number of pathways. This conclusion
stems from the fundamental property of a quantum phase: only a
gauge invariant phase is observable, while quantum phase
accumulated in a closed path is such an invariant. As for $N_p$
pathways, there exist $N_p -1$ closed paths. So it is implied that
multiple-pathway in FMO complexes is a necessary condition for the
enhancement of the EET by the quantum phases. It provides a strong
evidence to support the statement that the multiple energy
delivery pathway is also an acceptable contributing factor for
perfect energy transfer\cite{a1,a2}. Therefore,  we demonstrate
that, the quantum phases may bring the other two factors, the
optimal space of the pigments and multiple-pathway, together to
contribute the EET in photosynthetic complexes with perfect
efficiency.

The paper is organized as follows.  In Sec. II, we present the
complex quantum network we used to study the EET in photosynthetic
complexes. A particularly simple and illustrative example with
three sites of complex network is presented in Sec. III, where
some main conclusions, such as the phases may play an important
role in the EET and the optimized phases are insensitive to the
environments, are demonstrated in this very simple example. In
Sec. IV, we present a symmetric complex quantum network with $N_p$
pathways to show that the multi-pathway is also a contributing
factor for the perfect efficiency of the EET. In Sec. V, we
investigate the EET in FMO with our complex quantum network model.
Finally, the conclusions are presented in Sec. VI.

\section{Quantum network model with the phases}

The system we consider is a quantum network of $N$ connected sites
(nodes), schematically shown in Fig.1.
\begin{figure}[htbp]
\begin{center}
\vspace{-1.2cm}
\includegraphics[width=11cm]{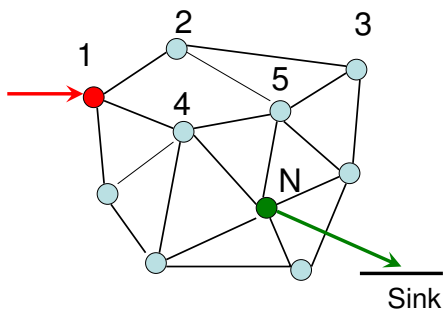}
\vspace{-3.5cm}
\caption{(Color online) The schematic representation of the
quantum network which is a collection of $N$ connected sites
(nodes). Each site is modeled as a spin-1/2 particle (qubit). The
particles are interacting with each other (solid lines) in the
quantum network and may suffer dissipative losses as well as
dephasing. An excitation is initialled at site $1$. The arrow
between site $N$ and sink denotes an irreversible transfer of
excitations from site $N$ to the sink.} \label{Fig0}
\end{center}
\end{figure}
Each site is modelled here as a spin-1/2 particle and it may
support an excitation which can be exchanged between lattice sites
by hopping. The initial (input) state is an excitation state which
describes an excitation localized at site $1$ (or several sites).
We are interested in the transfer rate that the excitation
transfers from the input state to the sink. The quantum evolution
of the network of $N$ sites is usually described by a Hamiltonian
of the form
\begin{equation}
\label{H}
H=\sum_{j=1}^{N}\epsilon_{j}\sigma_{j}^{+}\sigma_{j}^{-}+\sum_{j\neq
l}
V_{jl}(\sigma_{j}^{+}\sigma_{l}^{-}+\sigma_{l}^{+}\sigma_{j}^{-}),
\end{equation}
where  $\sigma_{j}^{+}$ and $\sigma_{j}^{-}$ are the raising and
lowering operators for site $j$.  $\sigma_{j}^{+}$=$|j \rangle
\langle 0 |$ and $\sigma_{j}^{-}$=$|0 \rangle \langle j|$, where
$|0\rangle$  represents the zero exciton state of the system and
$|j\rangle$ denotes the excitation at site $j$. The site
energy and two-body coupling strength are given by the real
numbers $\epsilon_{j}$ and $ V_{jl}$, respectively. The quantum
network\cite{Gnutzman} described in Eq.(\ref{H})
 has been used to study the EET in photosynthetic complexes in much literature \cite{b1,b2,b3,b4,b6,b7,b8,b9,b10,b11}.

In this paper, we add a quantum phase factor $e^{-i\phi_{jl}}$
with $\phi_{jl}$ a real number to the hopping term between sites
$j$ and $l$. The phase factor is determined by the detailed
structure of the quantum network. As for FMO, the phases are
related to the length of the pigments as well as the intrinsic
features of the barriers between the adjacent pigments. In this
case, the Hamiltonian (1) is replaced by
\begin{equation}
\label{H1}
H=\sum_{j=1}^{N}\epsilon_{j}\sigma_{j}^{+}\sigma_{j}^{-}+\sum_{j\neq
l}
V_{jl}(e^{-i\phi_{jl}}\sigma_{j}^{+}\sigma_{l}^{-}+e^{i\phi_{jl}}\sigma_{l}^{+}\sigma_{j}^{-}).
\end{equation}
Compared with the quantum network with real variables in
Eq(\ref{H}), this model can be named as a complex quantum network
model. We will show that the quantum phase factors play the
fundamental role in energy transfer of the photosynthetic
complexes.

As usual, we assume that all sites are susceptible simultaneously
to two distinct types of noise processes.  The first one is  a
dissipative process that transfers the excitation energy in site
$j$  to the environment with rate $\Gamma_{j}$, which leads to
energy loss.  The second one is a pure dephasing process with rate
$\gamma_{j}$  which destroys the phase coherence of any
superposition state in the system.    The dissipative and the pure
dephasing processes are described, respectively, by the Lindblad
super-operators \cite{b2,b3,b4},

\begin{equation}
L_{diss}(\rho)=\sum_{j=1}^{N}\Gamma_{j}[-\{ \sigma_{j}^{+}\sigma_{j}^{-} , \rho \} +2\sigma_{j}^{-}\rho\sigma_{j}^{+} ],
\end{equation}

\begin{equation}
L_{deph}(\rho)=\sum_{j=1}^{N}\gamma_{j}[-\{ \sigma_{j}^{+}\sigma_{j}^{-} , \rho \} +2\sigma_{j}^{+}\sigma_{j}^{-}\rho\sigma_{j}^{+} \sigma_{j}^{-}],
\end{equation}
where $\{A,B\}$ is an anticommutator. The absorption of the energy
from site $k$ to the sink (numbered $s$) is modeled by a
Lindblad operator
\begin{equation}
 L_{s}(\rho)=\Gamma_{s}[2\sigma_{s}^{+}\sigma_{k}^{-}\rho\sigma_{k}^{+}\sigma_{s}^{-}-\{\sigma_{k}^{+} \sigma_{s}^{-}\sigma_{s}^{+}\sigma_{k}^{-},\rho\}],
\end{equation}
where $\Gamma_{s}$ is the trapping rate. This term describes the
irreversible decay of the excitations to the sink. So the full
time evolution of the density matrix $\rho$ of the system is
described by the master equation
\begin{equation}
 \frac{d\rho}{dt}=-\frac{i}{\hbar}[H,\rho]+L_{diss}(\rho)+L_{deph}(\rho)+L_{s}(\rho).
 \end{equation}
\indent The efficiency of EET is measured by the population $P_{sink}$  transferred to the sink from the site $k$  \cite{b2,b3,b4},
\begin{equation}
P_{sink}=\rho_{sink}(\infty)=2\Gamma_{s}\int_{0}^{\infty}\rho_{kk}(t)dt.
\end{equation}

\section{Bipathway Quantum network}

To study the role of the phase in quantum network, a particularly
simple and illustrative example shown in Fig. 2 is to study
quantum transport in a system of three sites.
The exciton is transferred from site 1 to site 3 through two
pathways, and finally is trapped by the sink with rate
$\Gamma_{s}$. Sites 1, 2, and 3 are susceptible simultaneously
to the dissipative and the pure dephasing processes. The dynamics
of the system can be described by Eqs. (2-7). If we choose
$\epsilon_{1}=\epsilon_{2}=\epsilon_{3}=\epsilon$,
$V_{12}=V_{23}=V_{13}=V$,
$\Gamma_{1}=\Gamma_{2}=\Gamma_{3}=\Gamma$,
$\gamma_{1}=\gamma_{2}=\gamma_{3}=\gamma$, the analytical
expression of $P_{sink}$ (in the Appendix) can be obtained

\begin{widetext}
\begin{equation}\label{}
    P_{sink}=\frac{V^{4}\Gamma_{s}[A^{2}V^{4}\sin^{2}\phi-AV\Gamma(DB^{2}+AV^{2})\sin\phi+BD(D\Gamma B^{2}+ACV^{2})]}{A^{2}V^{6}
    (3\Gamma+\Gamma_{s})\sin^{2}\phi+G},
\end{equation}
\end{widetext}
where the phase difference $\phi=\phi_{12}+\phi_{23}-\phi_{13}$,
$G=(D\Gamma
B^{2}+ACV^{2})[D\Gamma(\Gamma+\Gamma_{s})B^{2}+C(2\Gamma\Gamma_{s}+
\gamma\Gamma_{s}+3D\Gamma)V^{2}]$, $A=3\Gamma+\Gamma_{s}+3\gamma$,
$B=2\Gamma+\Gamma_{s}+2\gamma$, $C=3\Gamma+\Gamma_{s}+2\gamma$,
and $D=\Gamma+\Gamma_{s}$. Note that the corresponding
analytical expression of $P_{sink}$ for the real coupling rates is
obtained in Ref. \cite{b1}. Although there are three phase
factors $\phi_{12}$, $\phi_{23}$, and $\phi_{13}$ in this
three-site network, it is notable that only the phase difference
$\phi=\phi_{12}+\phi_{23}-\phi_{13}$ is independent. It
demonstrates the fact that only the phase difference accumulated
in the two pathways ( pathway $1 \rightarrow 2 \rightarrow 3$
and pathway $1 \rightarrow 3 $ ) affects the interference at site
$3$.
\begin{figure}[htbp]
\begin{center}
\vspace{-1.5cm}
\includegraphics[width=11.5cm]{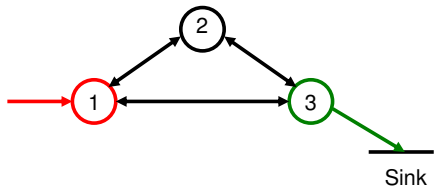}
\vspace{-4.5cm}
\caption{(Color online) The schematic representation of a bipathway quantum network. The exciton is
transferred from site 1 to site 3 through two pathways, and finally is trapped by the sink with the rate $\Gamma_{s}$.}
\label{fig2}
\end{center}
\end{figure}

From Eq. (8), we can easily find that $P_{sink}$ always increases
with $V$, while it always decreases with $\Gamma$.
 However, $P_{sink}$ is not a monotonic function of $\gamma$ and $\Gamma_s$. We plot the dependence of
  $P_{sink}$ on the different parameters in Fig.3.

\begin{figure}[htbp]
\begin{center}
\includegraphics[height=4cm]{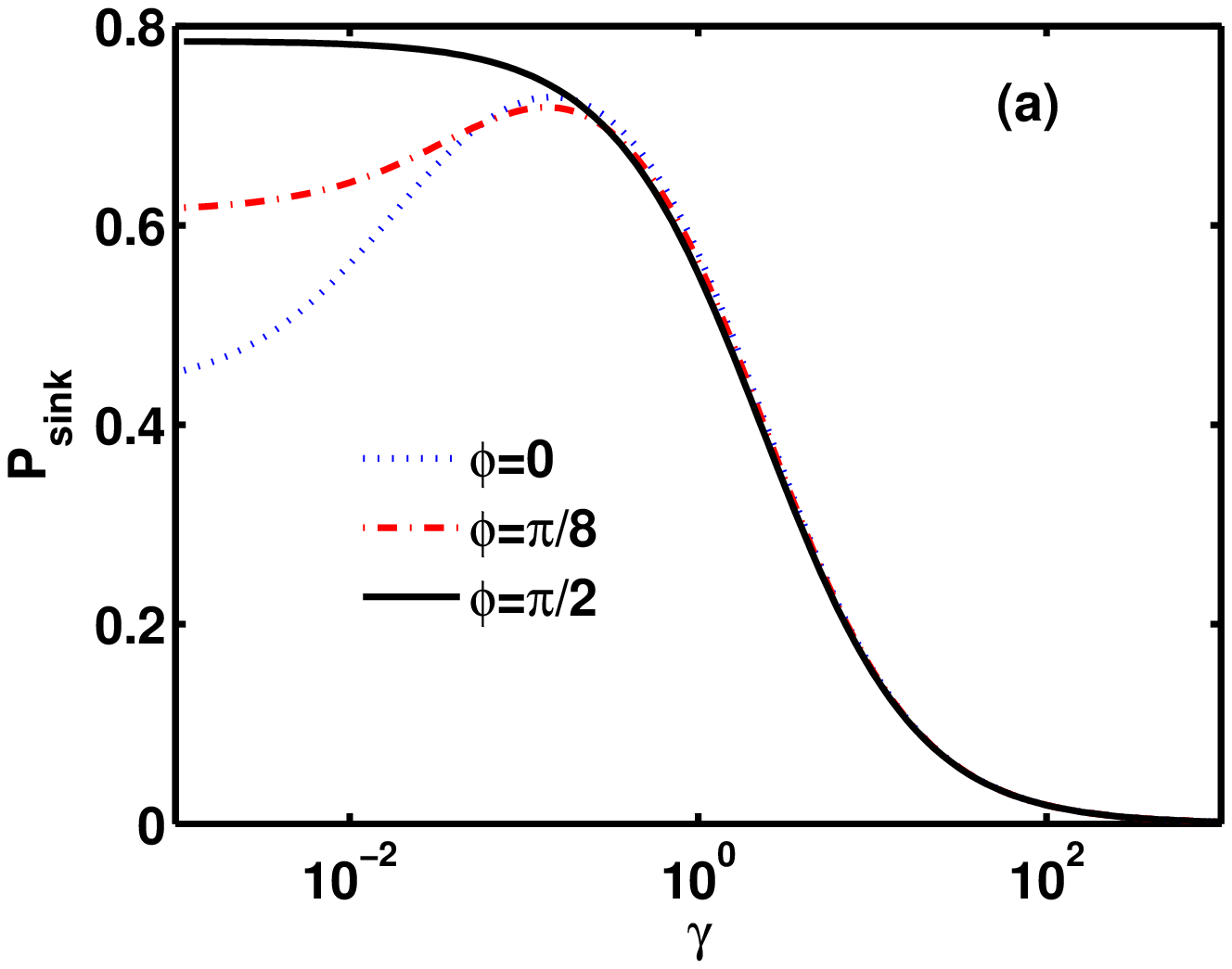}
\includegraphics[height=4cm]{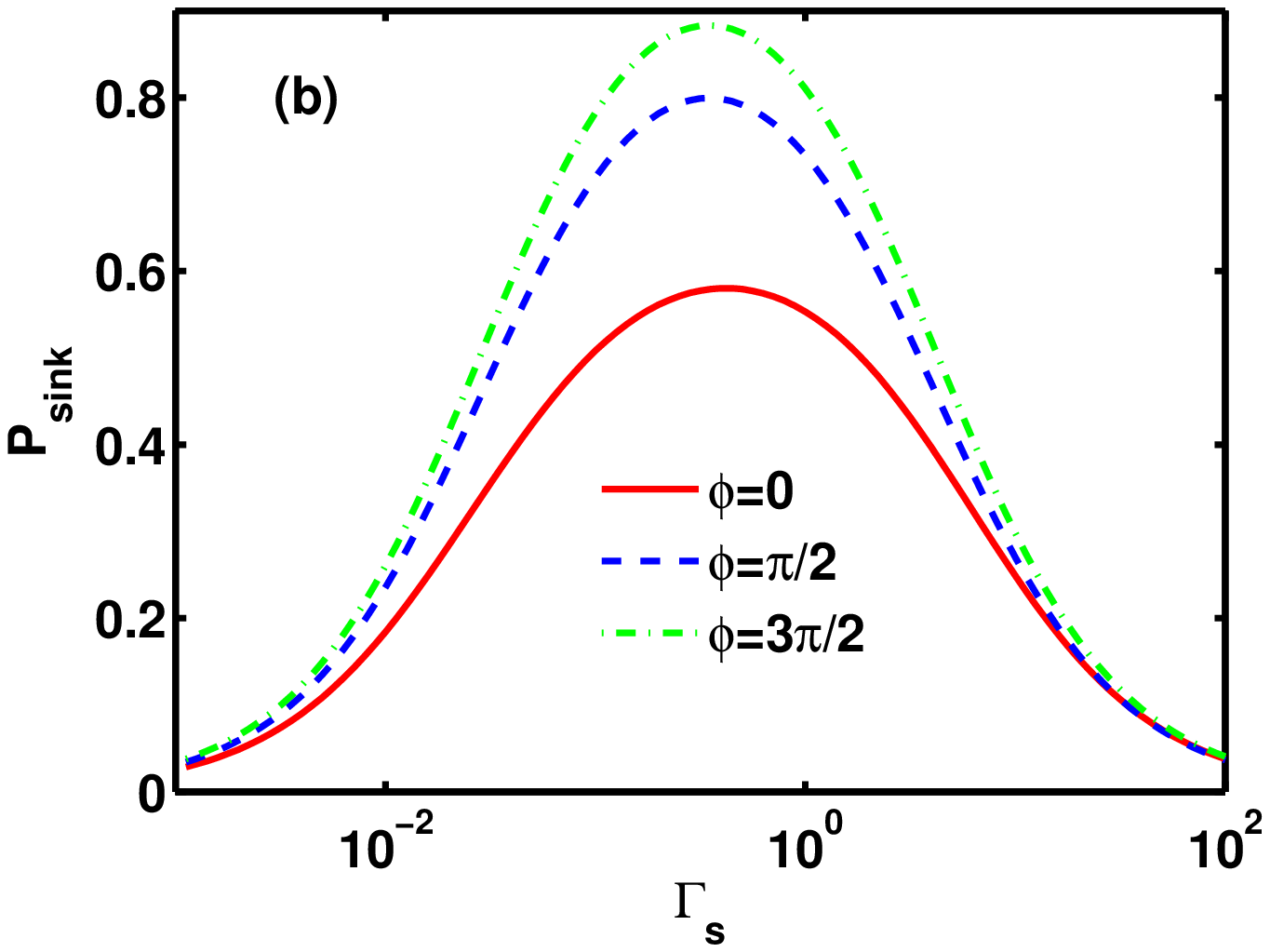}
\includegraphics[height=4cm]{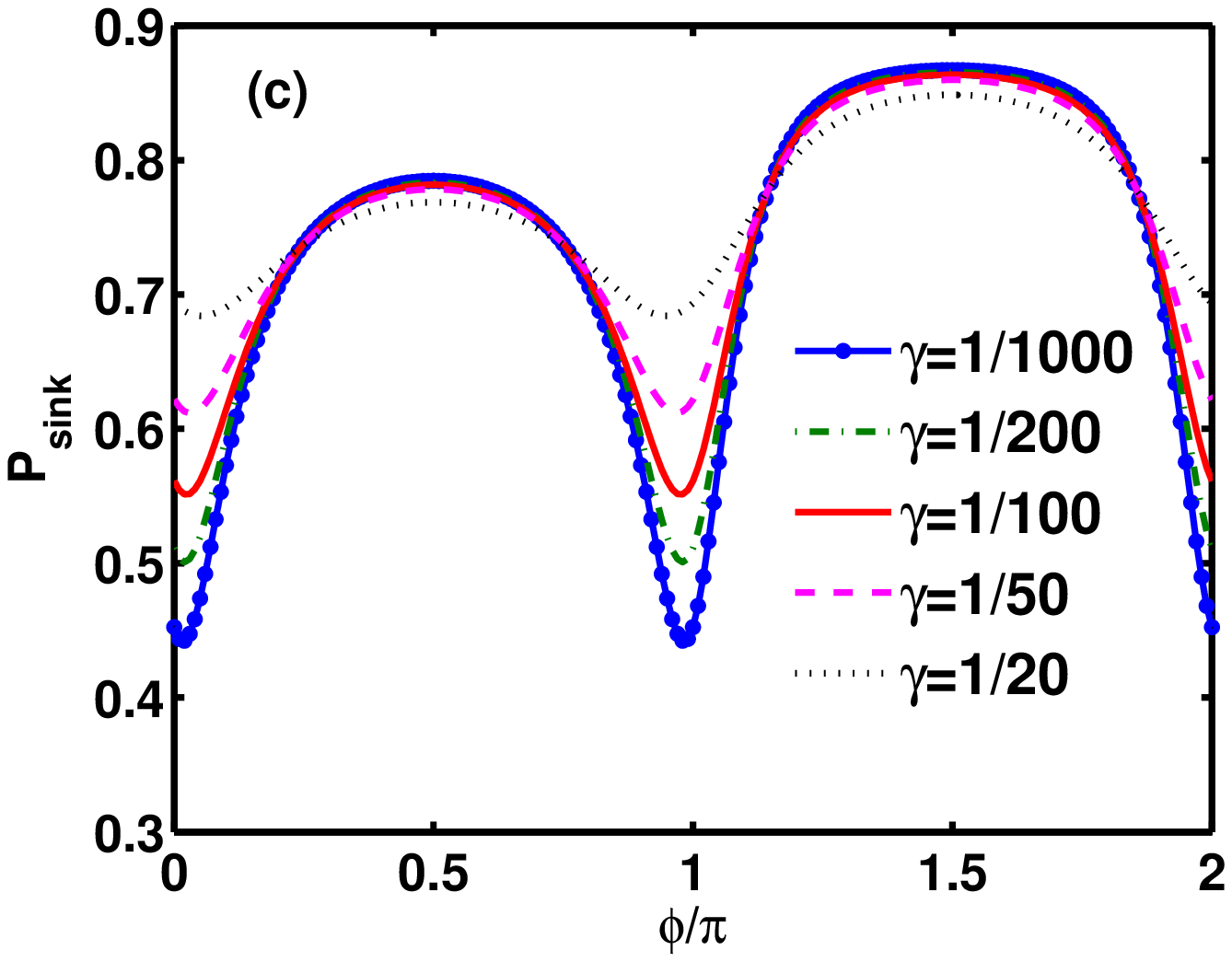}
\includegraphics[height=4cm]{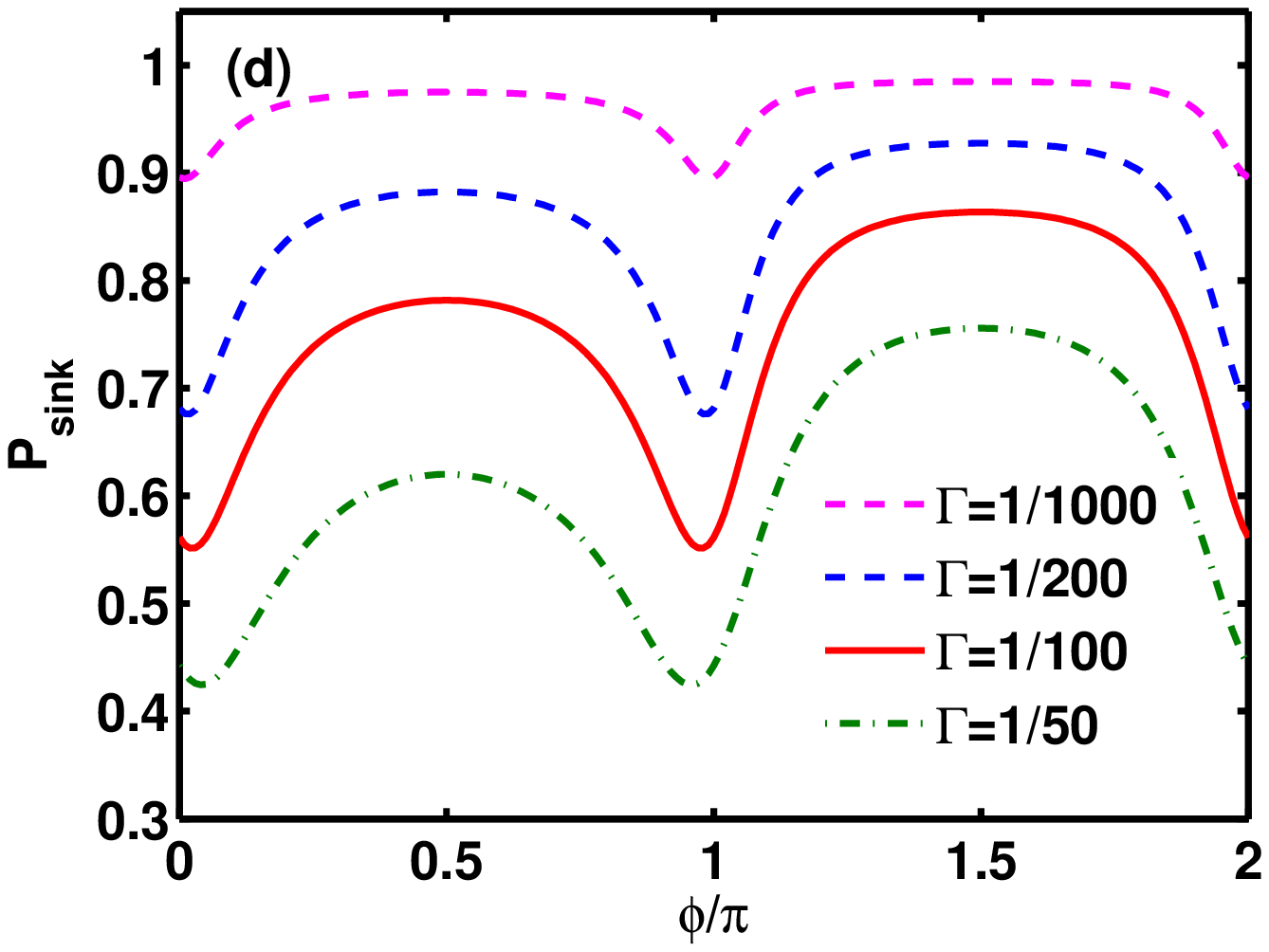}
\caption{(Color online) The dependence of the transfer efficiency
$P_{sink}$ on the different parameters. (a) $P_{sink}$ vs $\gamma$
for different values of $\phi$. (b) $P_{sink}$ vs $\Gamma_s$ for
different values of $\phi$.  (c) $P_{sink}$ vs $\phi$ for
different values of $\gamma$. (d) $P_{sink}$ vs $\phi$ for
different values of $\Gamma$. Unless otherwise noted, the
parameters are $V=1/5$, $\Gamma_s=1/5$, $\Gamma=1/100$, and
$\gamma=1/100$. }.
\end{center}
\end{figure}

Figure 3(a) shows the relation between $P_{sink}$ and $\gamma$ for
different values of $\phi$. For $\phi=0$, there exists an optimal
value of $\gamma$ at which $P_{sink}$ is maximal, which indicates
that the dephasing from the noise may even facilitate the EET.
Note that the similar conclusion is extensively reported in the
previous works \cite{b2,b3,b10}.  When the phase $\phi$ is
considered, the phase can change the efficiency remarkably at low
dephasing (purely quantum mechanical), while the efficiency is not
sensitive to the phase  at large dephasing (quantum coherent
destroyed). Therefore, the phase in EET plays a key role at low
dephasing.

Figure 3(b) shows the dependence of the transfer efficiency
$P_{sink}$ on the trapping rate $\Gamma_s$ for different values of
$\phi$. When $\Gamma_s$ is very small, the system couples weakly
to the sink, few exciton can reach the sink and the efficiency
tends to zero. When $\Gamma_s$ is too large, the trapping rate
$\Gamma_s$ mismatches the transport rate of the exciton in the
quantum network, thus the efficiency also goes to zero. Therefore,
there exists an optimal value of $\Gamma_s$ at which the
efficiency takes its maximal value.

Figures 3(c) and 3(d) show the efficiency $P_{sink}$ as a function
of the phase $\phi$ for different values of $\gamma$ and $\Gamma$,
respectively. It is found that there are two optimal values (about
$\pi/2$ and $3\pi/2$) of $\phi$ at which $P_{sink}$ takes its
extremum value, especially, it reaches a maximum value at
$3\pi/2$. The minimal value of $P_{sink}$ appears at $\phi \approx
0$ , $\pi$ and 2$\pi$. Obviously, the maximal values of $P_{sink}$
are due to the constructive interference, while its minimal values
are due to the destructive interference. Interestingly, the phases
which correspond to the extremal values of $P_{sink}$ almost do
not change with $\gamma$ and $\Gamma$.

\indent Therefore, we can conclude that quantum phase in the
two-body couplings plays a key role in the EET, especially at low
dephasing. Remarkably, the optimized phases in the quantum network are
almost independent of the environments ($\gamma$ and $\Gamma$).

\section{Multiple-pathway quantum network}

Since the number of pathways is an important quantity in
quantum network,  it is necessary to investigate the role of
multiple pathways on the efficiency of EET. For simplicity, we
consider a symmetric complex quantum network including $N_p$
pathways shown in Fig. 4.
\begin{figure}[htbp]
\begin{center}
\includegraphics[width=12.5cm]{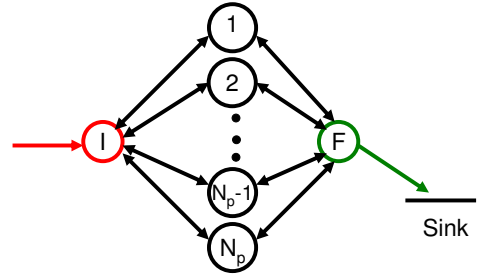}
\vspace{-4.5cm}
\caption{(Color online) The schematic representation of the
symmetric quantum network including $N_p$ pathways. An excitation
is initiated at site $I$ and transferred to site $F$ through
$N$ pathways. The arrow between site $F$ and sink denotes an
irreversible transfer of excitations from site $F$ to the sink.}
\label{Fig4}
\end{center}
\end{figure}
The exciton is transferred from site $I$ to site $F$ through
multiple pathways, and finally is trapped by the sink. $N_{p}$
describes the number of pathways between the sites $I$ and
$F$.  The dynamics of the system can also be described by Eqs.
(2-7). From Eqs. (2-7), we can  obtain the efficiency $P_{sink}$
for different number $N_p$ of pathways.
\begin{figure}[htbp]
\begin{center}
\includegraphics[height=5.0cm]{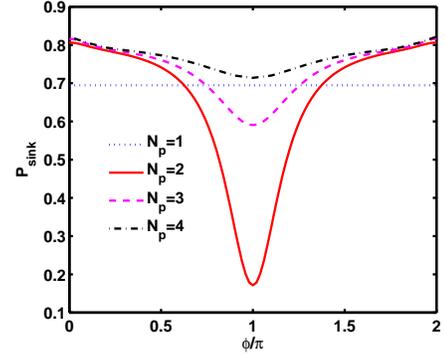}
\caption{(Color online) Transfer efficiency $P_{sink}$ as a
function of the phase difference $\phi$ for different number $N_p$
of the pathways. The other parameters are $\Gamma=1/100$,
$\gamma=1/100$, $\Gamma_{s}=1/5$, and $V=1/5$.} \label{default}
\end{center}
\end{figure}

Figure 5 shows the dependence of the transfer efficiency
$P_{sink}$ on the phase $\phi$ for different number $N_p$ of the
pathways. For a symmetric network, we choose
$\epsilon_{j}=\epsilon$, $V_{Ij}=V_{Fj}=V$,
$\Gamma_{j}=\Gamma$,$\gamma_{j}=\gamma$, $j=1,2,3...N_p$.  There
are $N_p-1$ independent phases because of $N_p-1$ closed loops in
the system. For simplicity, we only vary $\phi_{I1}$ ($=\phi$) and
the other phases are set to zero. For a single pathway ($N_p=1$),
the efficiency $P_{sink} $ is always equal to $0.695$, this is due
to the fact that no quantum interference can occur in a single
pathway. For double pathways ($N_p=2$), the quantum interference
at site $F$ occurs and the phases take effect. Due to the
destructive interference, there exists a minimal value of the
transfer efficiency at $\phi=\pi$. The efficiency $P_{sink}$ takes
its maximal value at $\phi=0$ or $ 2\pi$, where the constructive
interference occurs.  The efficiency of multiple pathways at
constructive interference has an enhancement compared with the
single pathway.  As the number of the pathways increases, the
effects of destructive interference on the efficiency decrease. We
here have assumed that all other phases in $N_p>3$ pathways are
zero. If we further optimize those phases, the enhancement by the
multiple-pathway is clearer. It supports the conclusion in
the quantum scattering model\cite{b16} where the resonance transport
is enhanced in multiple-pathway. Therefore, the multiple-pathway
can reduce the destructive interference and facilitate EET in the
quantum network. It seems that most local minima or maxima
occur at multiples of $\pi/2$, but the accumulated phases for
local minima or maxima are the multiples of $\pi/2$ only when all
pathways are the same. The quantum interference at the given site
is determined by the accumulated phase in the multiple pathways.
When all pathways are the same, the constructive interference and
the destructive interference occur at $\phi=0, \pi, 2\pi$,
respectively. However, when the pathways are not the same, the
interference in the closed pathways becomes complicated. The phase
conditions for the constructive interference and the destructive
interference depend on the system parameters, such as the site
energy, the coupling strength, and the number of the sites in each
pathway.

\indent Note that the similar findings are also found in
Ref.\cite{Cao}. They defined an effective hopping rate as the
leading order picture and
    nonlocal kinetic couplings as the quantum correction and found that the optimized multiple pathways can
    suppress the destructive interference in nonlinear network configurations.
    Although the model and the method are different from ours, the impact of closed paths on the transport are the same in nature.


\section{Excitation energy transfer in FMO complex}

The architecture of antenna light-harvesting complexes varies
widely among photosynthetic organisms. A well-studied example is
the water-soluble FMO complex of green sulfur bacteria. FMO
complex essentially acts as a molecular wire, transferring
excitation energy from the chlorosomes, which are the main
light-harvesting antennae of green sulfur bacteria, to the
membrane-embedded reaction center. The FMO is a trimer made of
three identical subunits, each containing seven
pigments\cite{Note}. Because the inter-subunit coupling is
vanishingly small, we only consider the dynamics of EET within
one subunit.  The subunit containing seven pigments shown in Fig.
6(a) can be modeled as a network of seven sites with site
dependent coupling and site energies. We use the experimental
Hamiltonian of FMO given in \cite{matrix}, and the matrix of the
Hamiltonian takes the form
\begin{widetext}
\begin{equation}
\label{H_FMO} H= \left(
\begin{array}{ccccccc}
 \mathbf{215} & \mathbf{-104.1} &5.1 &-4.3&4.7&\mathbf{-15.1}&-7.8\\
\mathbf{ -104.1}&\mathbf{220.0}&\mathbf{32.6}  &7.1&5.4&8.3&0.8\\
 5.1&\mathbf{32.6} &\mathbf{0.0} &\mathbf{-46.8}&1.0&-8.1&5.1\\
 -4.3&7.1 &\mathbf{-46.8} &\mathbf{125.0}&\mathbf{-70.7}&-14.7&\mathbf{-61.5} \\
 4.7 & 5.4&1.0 &\mathbf{70.7}&\mathbf{450}&\mathbf{89.7}&-2.5\\
 \mathbf{-15.1} & 8.3 &-8.1 &-14.7&\mathbf{89.7}&\mathbf{330.0}&\mathbf{32.7}\\
 -7.8& 0.8 &5.1 &\mathbf{-61.5}&-2.5&\mathbf{32.7}&\mathbf{280}
\end{array}
\right)
\end{equation}
\end{widetext} with units of cm$^{-1}$ and a total offset of
12230cm$^{-1}$ to set the lowest site energy to zero for
convenience (This overall shift in energy does not affect the
dynamics of the system).  In units with $\hbar=1$, we note that
the rate  $1 ps^{-1}\equiv5.3 $ cm$^{-1}$. By neglecting the
couplings weaker than $15$ $cm^{-1}$ (only bold entries in the
Hamiltonian are considered) in this model Hamiltonian, the
transport in an individual monomer of FMO can be mapped to a
quantum network shown in Fig. 6(b).
\begin{figure}[htbp]
\begin{center}
\includegraphics[width=8cm]{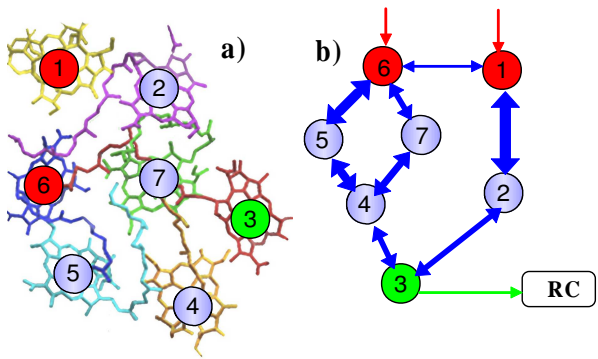}
\caption{(Color online)(a) The spatial structure of one monomeric
subunit of the FMO complex \cite{a7,b10}. Each monomer has seven
pigments labeled by $\textcircled1-\textcircled7$. The initial
state is taken to be a superposition state located at pigments
1 and 6. and pigment 3 is in the vicinity of the reaction center
(RC). (b) The simplified network for the monomeric subunit of
FMO. The thickness of two-headed arrow indicates the coupling
strengths and only couplings above 15 cm$^{-1}$ are shown. The
exciton is transferred from sites 1 and 6 to site 3 through the
network, and finally trapped by the reaction center with the rate
$\Gamma_{s}$.} \label{default}
\end{center}
\end{figure}

However, the Hamiltonian in Eq.(\ref{H_FMO}) may not be sufficient
to describe the EET in the FMO. We here focus on the possible
effects of the newly added phase factors in the coupling terms.
From Eq. (\ref{H1}) and Fig. 6(b) we can find that there are eight
phases, $\phi_{12}$, $\phi_{23}$, $\phi_{34}$, $\phi_{45}$,
$\phi_{47}$, $\phi_{56}$, $\phi_{67}$, $\phi_{16}$, but only two
phase differences,
$\phi_1=\phi_{61}+\phi_{12}+\phi_{23}-\phi_{67}-\phi_{74}-\phi_{43}$
and $\phi_2=\phi_{67}+\phi_{74}-\phi_{65}-\phi_{54}$, which are
independent since there are just two independent closed loops.
Therefore,  without loss of the generality, we vary the phases
$\phi_{12}$ and $\phi_{67}$ and the other phases are set to zero
in our numerical simulations.

The initial state for our simulation  is a superposition state
localized at pigments 1 and 6 which are close to the chlorosome
antenna (donor). It can be written as
$|\Psi(0)\rangle=\alpha|1\rangle+\beta|6\rangle$ with
$|\alpha|^{2}+|\beta|^{2}=1$.  Pigment 3 is the main
excitation donor to the reaction center. The energy trapping rate
from pigment 3 to the center in the literature \cite{b2,b3,b4}
ranges from 1 ps$^{-1}$ to 4 ps$^{-1}$.  In our calculations,
 we chose $\Gamma_{s}=20/1.88$ cm$^{-1}$ corresponding to about 2ps$^{-1}$.   The measured lifetime of excitons  is of the order of
 1ns which determines a dissipative decay rate of $0.5/188$cm$^{-1}$. Unless otherwise noted, we choose $ \Gamma=0.5/188$cm$^{-1}$
 and $\gamma=0.01\Gamma$ in this paper and assume that $\Gamma$ and $\gamma$ are the same for each site.
 From Eqs. (2-7), we can numerically obtain the efficiency $P_{sink}$ of the EET in FMO complex for different cases.

\begin{figure}[htbp]
\begin{center}
\includegraphics[height=4cm]{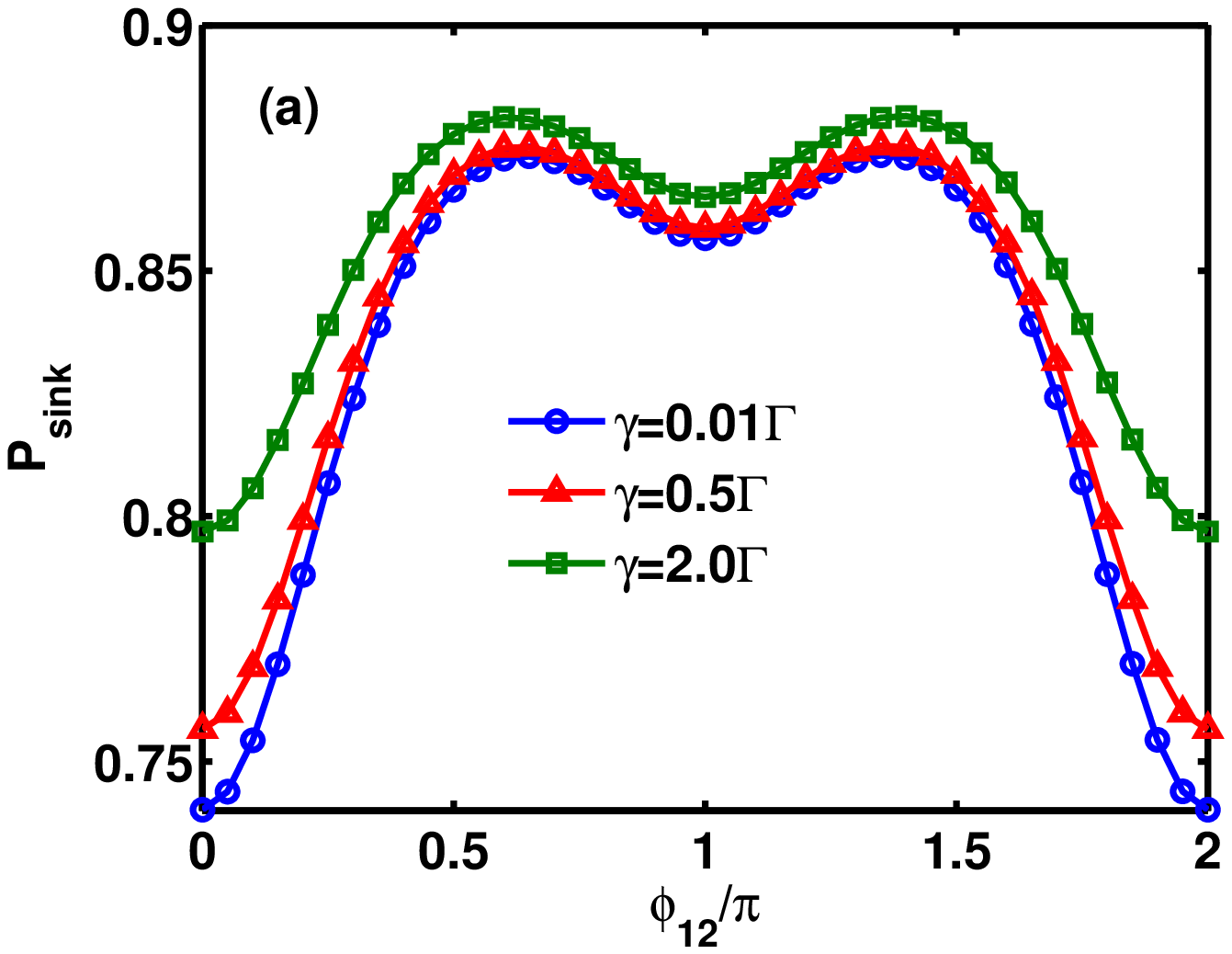}
\includegraphics[height=4cm]{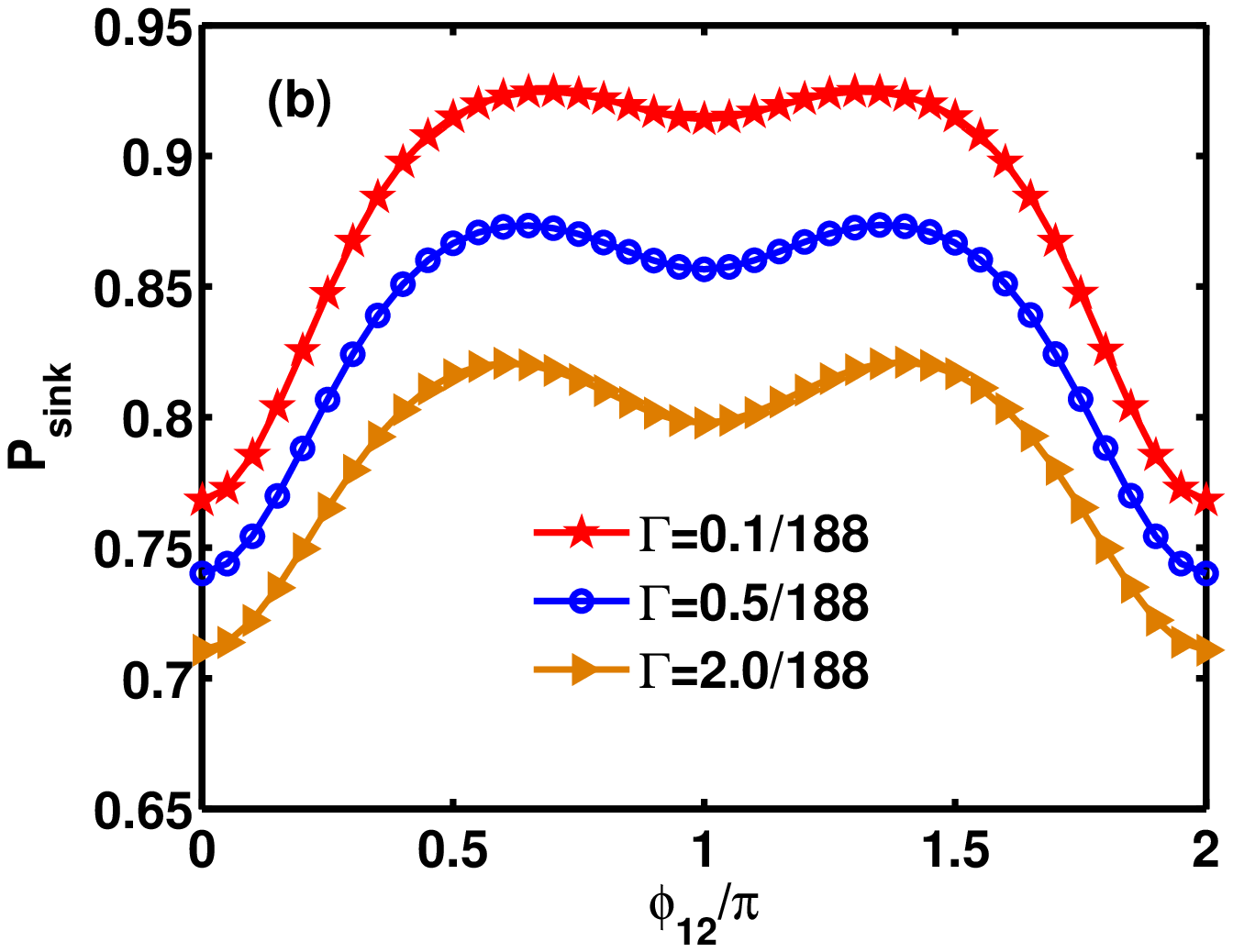}
\caption{(Color online) Transfer efficiency $P_{sink}$ as a
function of the phase $\phi_{12}$. (a)For different dephasing
rates $\gamma$ at $\Gamma=0.5/188$ cm $^{-1}$. (b) For different
dissipative rates $\Gamma$ at $\gamma=0.5/18800$ cm$^{-1}$.
 The other parameters are $\Gamma_{s}=20/1.88$ cm$^{-1}$, $\phi_{67}=0$ and $\alpha=\beta=\frac{\sqrt{2}}{2}$.}
\label{default}
\end{center}
\end{figure}

\indent Figure 7(a) and (b) shows the phase dependent efficiency
for different dephasing and dissipative rates with $\phi_{67}=0$
and $\alpha=\beta=\frac{\sqrt{2}}{2}$. It is found that there
exist two optimal values of $\phi_{12}$ (e. g.
$\phi_{12}\approx\pi/2$ or $3\pi/2$) at which the transfer
efficiency $P_{sink}$ takes its maximal value. When the dephasing
rate $\gamma$ or dissipative rate $\Gamma$ varies, the shape of
the curve in Fig. 7 almost does not change, which indicates an
important feature that the optimized phases are not sensitive to
the environment.  The phases denote the distance between the
pigments and the barriers between pigments, which are determined
actually by the spatial distribution of seven pigments. When the
pigments are optimally spaced, the exciton can pass through
FMO with optimal efficiency. Therefore, the phases from the
two-body interactions play a key role in energy transfer of the
FMO complex.

\begin{figure}[htbp]
\begin{center}
\vspace{-4.5cm}
\includegraphics[height=12cm]{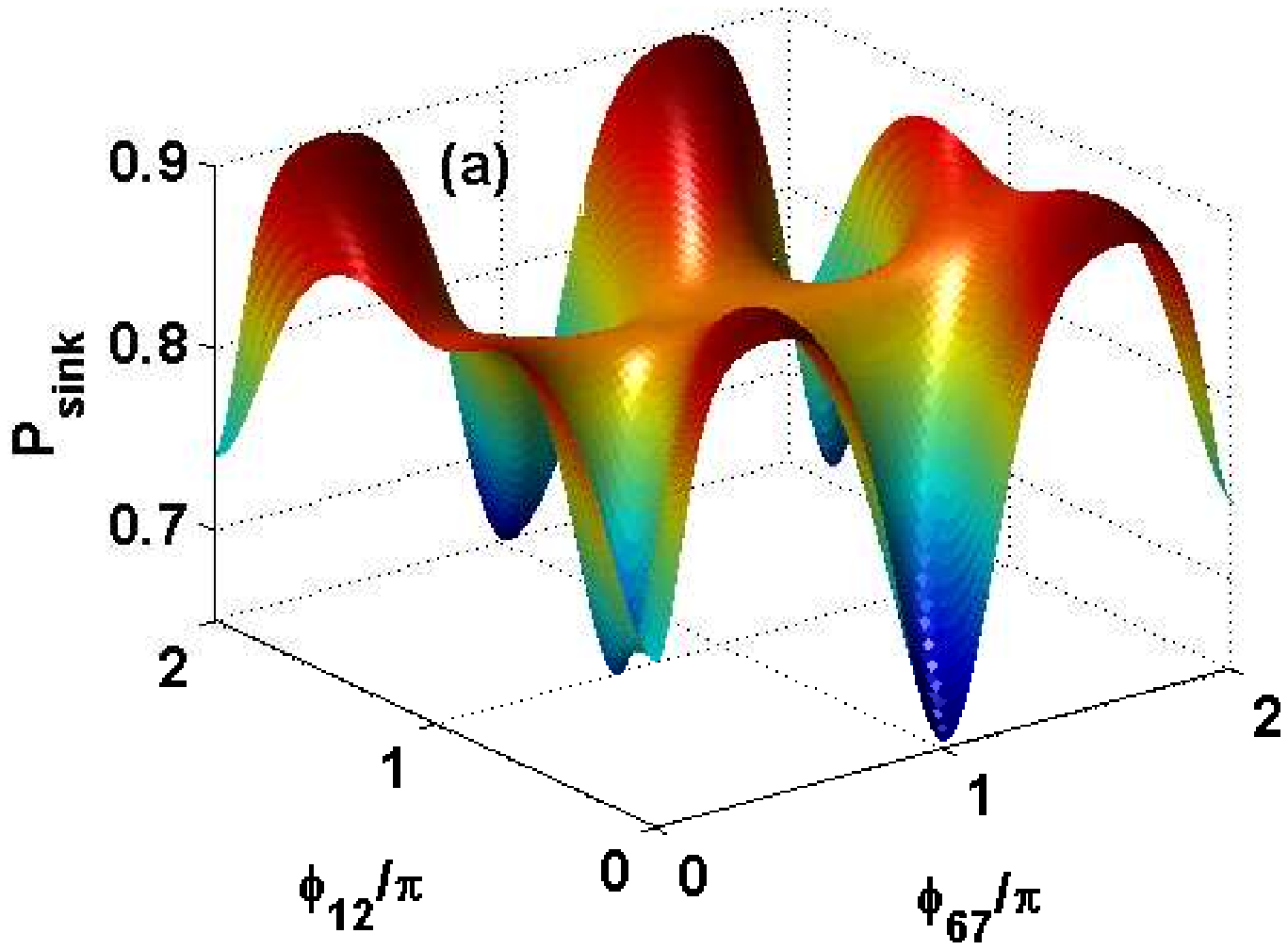}
\vspace{-3.5cm}
\vspace{-3.5cm}
\includegraphics[height=12cm]{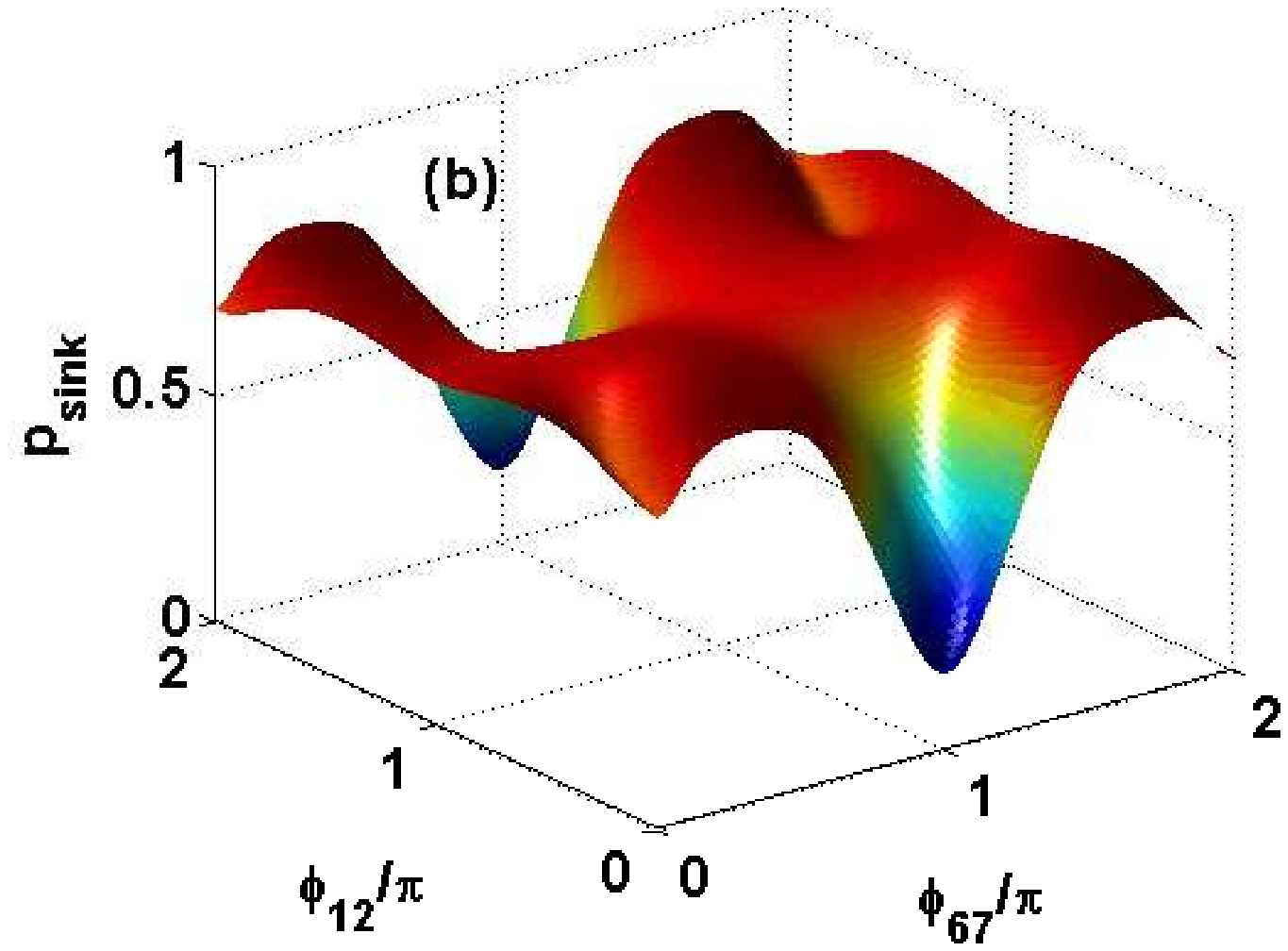}
\vspace{-3.5cm}
\vspace{-3.5cm}
\includegraphics[height=12cm]{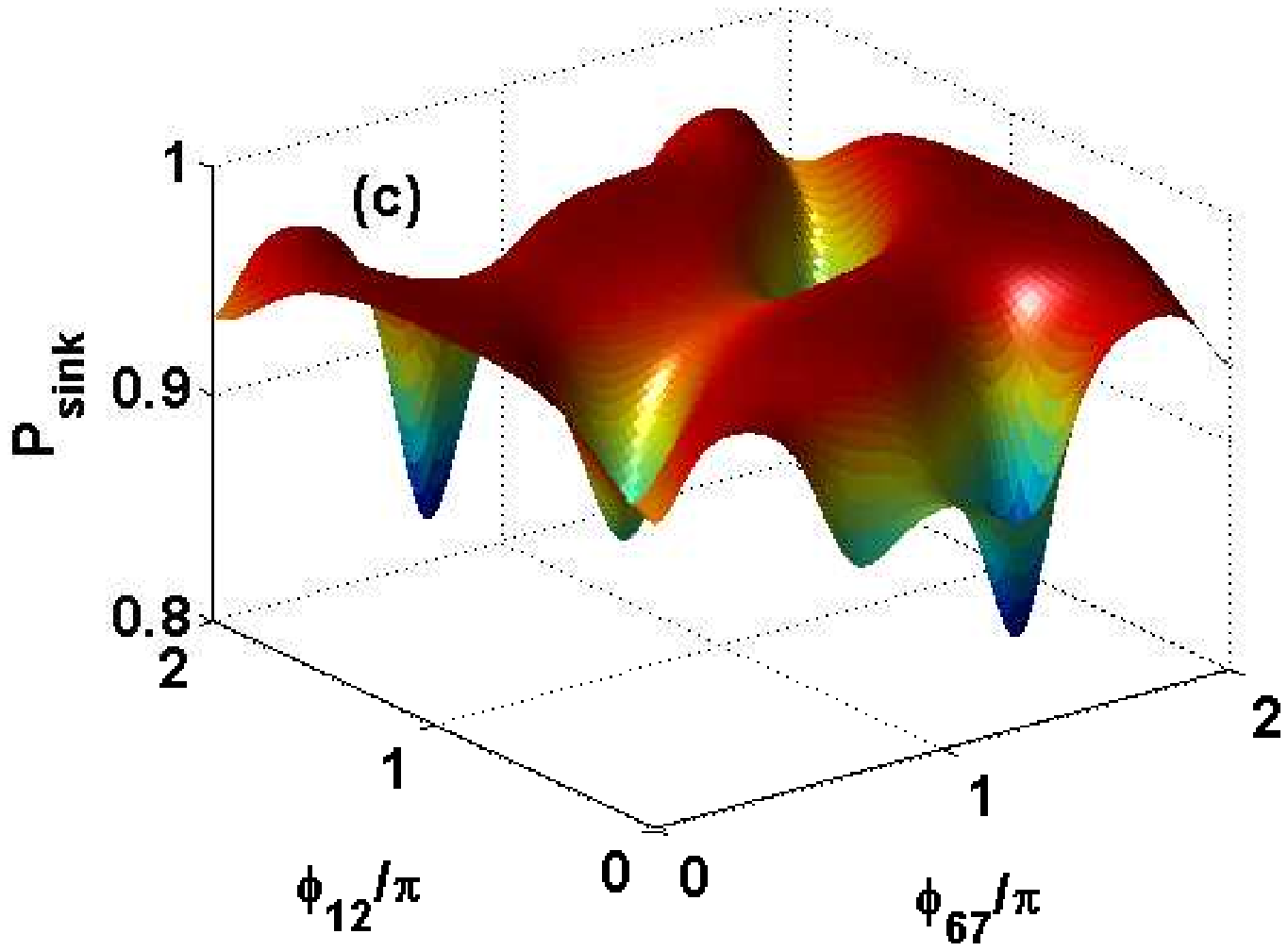}
\vspace{-3.5cm}
\caption{(Color online) Transfer efficiency $P_{sink}$ as a
function of the phases $\phi_{12}$ and $\phi_{67}$ for different
initial states. (a)$\alpha=\beta=\frac{\sqrt{2}}{2}$.
(b)$\alpha=0$ and $\beta=1$. (c)$\alpha=1$ and $\beta=0$. The
other parameters are $\Gamma_{s}=20/1.88$ cm$^{-1}$,
$\gamma=0.5/18800$ cm$^{-1}$ , and
 $\Gamma=0.5/188$ cm$^{-1}$. }
\label{default}
\end{center}
\end{figure}

Figure 8 shows the dependence of the transfer efficiency
$P_{sink}$ on the phases $\phi_{12}$ and $\phi_{67}$ for different
initial states.  To study the significance of the phases, we can
define the difference $\Delta P$ between the maximal
efficiency $P_{sink}^{max}$ and the minimal efficiency
$P_{sink}^{min}$, $\Delta P= P_{sink}^{max}- P_{sink}^{min}$. We
find that $\Delta P=0.2467$ for Fig. 8 (a)
($\alpha=\beta=\frac{\sqrt{2}}{2}$), $\Delta P=0.6459$ for Fig. 8
(b) ($\alpha=0$, $\beta=1$)  and $\Delta P=0.1432$ for Fig. 8 (c)
($\alpha=1$, $\beta=0$). Obviously, the phase can cause a
significant change in the efficiency and the change just slightly
depends on the initial states. The role of the phase coherence is
to overcome local energetic traps and aid efficient trapping
exciton energy by the pigments facing the reaction center. In this
case we can still find that there exist some optimal phase
regions where the transfer efficiency takes its maxima. The
optimal phase regions only slightly vary with the initial states.
Therefore, we demonstrate that the phase plays a significant role
in the EET and the optimal phase can facilitate the energy
transfer in THE FMO complex.

\begin{figure}[htbp]
\begin{center}
\includegraphics[height=5.0cm]{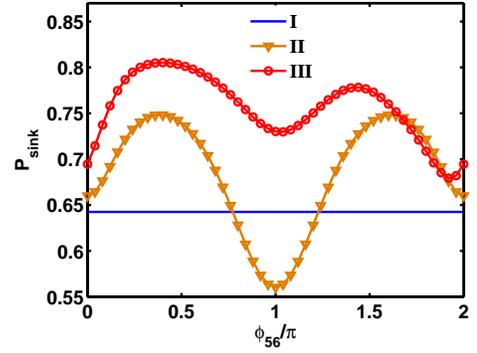}
\caption{(Color online) Transfer efficiency $P_{sink}$ as a function of the phase $\phi_{56}$
for different cases at $\alpha=0$ and $\beta=1$. (I)
One pathway: $\textcircled6$$\rightarrow$$\textcircled5$$\rightarrow$$\textcircled4$$\rightarrow$$\textcircled3$,
 namely,$V_{12}=V_{23}=V_{16}=V_{67}=V_{47}=0$. (II) Two pathways:
 $\textcircled6$$\rightarrow$$\textcircled5$$\rightarrow$$\textcircled4$$\rightarrow$$\textcircled3$
 and $\textcircled6$$\rightarrow$$\textcircled7$$\rightarrow$$\textcircled4$$\rightarrow$$\textcircled3$,
 namely, $V_{12}=V_{23}=V_{16}=0$. (III) Three pathways: the full quantum network of FMO
 with $\phi_{16}=0.1\pi$. $\Gamma_{s}=20/1.88$ cm$^{-1}$,
$\gamma=0.5/18800$ cm$^{-1}$ , and
 $\Gamma=0.5/188$ cm$^{-1}$. The other phases are set to zero.} \label{default}
\end{center}
\end{figure}

The experimental evidences \cite{a1,a2} show that besides the
optimal space distribution of the pigments, the multiple energy
delivery pathway is another acceptable contributing factor for
perfect energy transfer.  Therefore, it is necessary to study the
role of the multiple pathways in our model. In Fig. 9, the
efficiency of one pathway is compared with that of the multiple
pathways. It is found that the transfer efficiency $P_{sink}$ gets
a significant enhancement when the number of the pathways
increases. The efficiency is only $0.6425$ for a single pathway,
while it can reach $0.8$ for multiple pathways. Remarkably, the
efficiency in the full FMO (case III) even at destructive
interference is larger than that in a single pathway (case I).
This suggests that the dephasing noise in the system destroys the destructive interference and thus opens both paths for the transport \cite{b1,b3}.
This result supports the conclusion in the quantum scattering
model\cite{b16} where the resonance transport is enhanced in
multiple-pathways.  Therefore, we can conclude that multiple
pathways can also facilitate EET in the FMO complex which agrees with
the experimental statement \cite{a1,a2}.


 \section{Concluding remarks} In this paper, we have investigated the efficiency of the EET in
FMO complexes by adding the quantum phase factors to the quantum
network model. The phase describes the length of the pigments, the
distance, and the barriers between pigments and is then determined
by the space distribution of the pigments. We found that the
optimal distribution of the pigments can lead to the high
efficiency of the EET. Moreover, the optimal phase is not
sensitive to the environments. If the distribution of the pigments
is optimized, the efficiency always takes its maximal value, which
is indeed significant for high transfer efficiency. As we know,
the biological system governed by Darwinian selection has the
optimal structure, which can ensure that the quantum coherence
occurs in the optimal spatial distribution. In addition, we also
find that multiple pathways can facilitate EET in the FMO complex.
Therefore, we can conclude from the studies of the complex quantum
network model that, the optimal space distribution of the
pigments, the multitude of energy delivery pathways and the
quantum effects, are combined together to contribute to the
perfect energy transport in  FMO complexes.

In this paper we just add the phase factors phenomenally
to the two-body couplings in the Hamiltonian (9). Qualitatively,
the quantum phases are determined by the spatial structure of the
pigments in photosynthetic complexes; however, how to determine
them quantitatively in a microscopic theory or from the
experimental measurements is an important open question which
deserves further study.

Though we only have studied the transport process through the FMO
protein, the methods and conclusions can be extended to other
photosynthetic light-harvesting complexes. Furthermore,
understanding the mechanism of efficient energy transfer in
natural light-harvesting systems can help develop low-cost and
highly efficient man-made solar energy apparatuses, including
photovoltaic devices and artificial photosynthesis.

\indent This work was supported by the NNSFC (Nos.11175067, and 11125417),
the PCSIRT,
the SKPBRC
(No.2011CB922104), and the NSF of Guangdong (No.S2011010003323).

{\sl Note added. -- Shortly after we submitted the paper, there
was a preprint pasted in arXiv\cite{Yi2012}, where the effects of
the complex coupling are discussed.}

\appendix

\section{The derivation of Eq. (8)}
\indent In terms of the density matrix elements in the site basis $\rho_{ij}(t)$, the equations of motion for $N=3$ are
\begin{widetext}
\begin{equation}\label{}
    \frac{d\rho_{ij}}{dt}=-[2\Gamma+\Gamma_{s}(\delta_{iN}+\delta_{jN}+2\gamma-2\gamma\delta_{ij})]\rho_{ij}+iV[\sum_{l\neq j}e^{-i\phi_{il}}\rho_{il}-\sum_{l\neq i}e^{i\phi_{lj}}\rho_{lj}],
\end{equation}
\end{widetext}
\begin{equation}\label{}
    \frac{d\rho_{sink}}{dt}=2\Gamma_{s}\rho_{33}.
\end{equation}
\indent Since the exciton is transferred from site 1 to site 3, the initial conditions are
\begin{equation}\label{}
    \rho_{11}(0)=1, \rho_{sink}(0)=0,
\end{equation}
and the other density matrix elements $\rho_{ij}=0$.

\indent The system of coupled differential equations can converted into a set of algebraic equations via the Laplace transform. The above equations can be rewritten by the following
set of equations for the Laplace s-domain variables, for $\widetilde{\rho}_{11}$
\begin{widetext}
\begin{equation}\label{}
    (s+2\Gamma)\widetilde{\rho}_{11}-1-iV[\sum_{l\neq j}e^{-i\phi_{il}}\widetilde{\rho}_{il}-\sum_{l\neq i}e^{i\phi_{lj}}\widetilde{\rho}_{lj}]=0,
\end{equation}
\end{widetext}
and the other density matrix elements are
\begin{widetext}
\begin{equation}\label{}
    [s+2\Gamma+\Gamma_{s}(\delta_{iN}+\delta_{jN}+2\gamma-2\gamma\delta_{ij})]\widetilde{\rho}_{ij}-iV[\sum_{l\neq j}e^{-i\phi_{il}}\widetilde{\rho}_{il}-\sum_{l\neq i}e^{i\phi_{lj}}\widetilde{\rho}_{lj}]=0,
\end{equation}
\end{widetext}
\begin{equation}\label{}
    s\widetilde{\rho}_{sink}(s)-2\Gamma_{s}\widetilde{\rho}_{33}=0.
\end{equation}

\indent From Eqs. (A4-A6), we can easily obtain the expression of $\widetilde{\rho}_{sink}(s)$. From the relation of the Laplace transform for $t$ and $s$, we can find that
\begin{equation}\label{}
    P_{sink}=\rho_{sink}(\infty)=\lim_{s\rightarrow0}s\widetilde{\rho}_{sink}(s),
\end{equation}
and then the Eq. (8) is obtained.


\begin{thebibliography}{99}
\bibitem{a0}G. D. Scholes, G. R. Fleming, A. Olaya-Castro and R. V. Grondelle, Nature Chemistry\textbf{ 3}, 763 (2011);
Y. C. Cheng and G. R. Fleming, Annu. Rev. Phys. Chem. \textbf{60}, 241 (2009).
\bibitem{a1} R. V. Grondelle  and V. I. Novoderezhkin, Nature (London) {\bf 463}, 614
(2010);R. V. Grondelle,  J. P. Dekker, T. Gillbro, and V. Sundstrom, Biochim. Biophys. Acta {\bf1187}, 1 (1994).
\bibitem{a2}M. K. Sener, J. D. Olsen, C. N. Hunter, and K. Schulten,
Proc. Natl. Acad. Sci. USA {\bf 104}, 15723 (2007).
\bibitem{a3}G. S. Engel, T. R. Calhoun, E. L. Read, T. K. Ahn, T. Manal, Y. C. Cheng, R. E. Blankenship, and G. R.
Fleming, Nature (London) {\bf 446}, 782 (2007).
\bibitem{a4}H. Lee, Y. C. Cheng,  and G. R. Fleming, Science {\bf 316}, 1462 (2007).

\bibitem{a6} E. Collini, C. Y. Wong, K. E. Wilk, P. M. G. Curmi, P. Brumer,  and G. D. Scholes,
Nature (London) {\bf 463}, 644 (2010).


\bibitem{a5}G. Panitchayangkoona, D. Hayesa, K. A. Fransteda, J. R. Carama, E. Harela, J. Wenb,
R. E. Blankenshipb, and G. S. Engel, Proc. Natl. Acad. Sci. USA
{\bf 107}, 12766 (2010).




\bibitem{b1}M. B. Plenio and S. F. Huelga, New J. Phys. \textbf{10}, 113019 (2008); M. Mohseni, P. Rebentrost, S. Lloyd and A. Aspuru-Guzik,  J. Chem. Phys.\textbf{ 129}, 174106 (2008).
\bibitem{b2} S. Hoyer, M. Sarovar, and K. B. Whaley, New J. Phys. \textbf{12}, 065041(2010);
F. Fassioli and A. Olaya-Castro, New J. Phys.\textbf{ 12}, 085006(2010).
\bibitem{b3}F. Caruso, A. W. Chin, A. Datta, S. F. Huelga and M. B. Plenio, J. Chem. Phys. \textbf{131}, 105106 (2009); F. Caruso, S. Montangero, T. Calarco, S. F. Huelga, M. B. Plenio, Phys. Rev. A \textbf{85}, 042331 (2012).

\bibitem{b4}A. W. Chin, A. Datta, F. Caruso, S. F. Huelga and M. B. Plenio, New J. Phys. \textbf{12}, 065002 (2010).

\bibitem{b6}G. Panitchayangkoon, D. Hayes, K. A. Fransted, J. R. Caram, E. Harel, J. Wen, R. E. Blankenship and G. S. Engel,
Proc. Natl. Acad. Sci. USA \textbf{107}, 12766 (2010).
\bibitem{b7}Y. C.  Cheng and R. J. Silbey, Phys. Rev. Lett. \textbf{96},  028103 (2006).
\bibitem{b8}A. Nazir Phys. Rev. Lett. \textbf{103}, 146404 (2009).
\bibitem{b9} T. R. Calhoun, N. S. Ginsberg, G. S. Schlau-Cohen, Y. C. Cheng, M. Ballottari, R. Bassi and G. R. Fleming, J. Phys. Chem. B
 \textbf{113}, 16291(2009).


\bibitem{b10}P. Rebentrost, M. Mohseni, I. Kassal, S. Lloyd and A. Aspuru-Guzik, New J. Phys. \textbf{11}, 033003 (2009); P. Rebentrost,
M. Mohseni and A. Aspuru-Guzik,  J. Phys. Chem. B \textbf{113}, 9942(2009).
\bibitem{b11}L. James, R. Junghee,  L. Changhyoup, Y. Seokwon, J. Hyunseok and L. Jinhyoung, New J. Phys. \textbf{13}, 103002 (2011).
\bibitem{yi}B. Cui, X. X. Yi, C. H. Oh, J. Phys. B: At. Mol. Opt. Phys. 45, 085501 (2012); B. Cui, X. Y. Zhang, X. X. Yi, arXiv:1106.4429.

\bibitem{Ringsmuth} A. K. Ringsmuth, G. J. Milburn, and T. M. Stace, Nature Physics \textbf{8}, 562 (2012).
\bibitem{Scholak}  T. Scholak, F. Melo, T. Wellens, F. Mintert, and A. Buchleitner, Phys. Rev. E \textbf{83}, 021912 (2011).

\bibitem{a7}M. Sarovar, A. Ishizaki, G. R. Fleming, and K. Birgitta
Whaley, Nature Physics {\bf 6}, 462 (2010); A. Ishizaki and G. R.
Fleming, Proc. Natl. Acad. Sci. USA {\bf 106}, 17255(2009).

\bibitem{b5}A. Ishizaki and G. R. Fleming, J. Chem. Phys. \textbf{130}, 234111 (2009).

\bibitem{b12}S. Yang, D. Z. Xu, Z. Song, and C. P. Sun, J. Chem. Phys. \textbf{132}, 234501 (2011); J. Q. Liao,  J. F. Huang,  L. M. Kuang,  C. P. Sun, Phys. Rev. A \textbf{82}, 052109 (2010); H. Dong, D.-Z. Xu,
J.-F. Huang, and C.-P. Sun, Light: Sci. Appl. 1, e2 (2012).

\bibitem{b13}K. G. Pulak, Y. S.  Anatoly, and N. Franco, J. Chem. Phys. \textbf{134}, 244103 (2011)
\bibitem{b14}J. S. Cao,  J. Chem. Phys. \textbf{107}, 3204 (1997); J. L. Wu, F. Liu, Y. Shen, J. S. Cao,  R. J.  Silbey,
New J. Phys. \textbf{12}, 105012 (2010); J. Ye, K. Sun, Y. Zhao, Y. Yu, C. K. Lee, J. S. Cao, J. Chem. Phys. \textbf{136}, 245104 (2012).
\bibitem{baa}J. Prior, A.W. Chin, S. F. Huelga, M. B. Plenio, Phys. Rev. Lett. 105, 050404 (2010).
\bibitem{b15}X. T. Liang, Phys. Rev. E \textbf{82}, 051918 (2010) ; X. T. Liang, W. M. Zhang, and Y. Z. Zhuo, Phys. Rev. E \textbf{81}, 011906 (2010).

\bibitem{b16}B. Q. Ai and S. L. Zhu,  arXiv:1201.1740.

\bibitem{Jing} Y. Y. Jing,  R. H. Zheng, H. X. Li, and Q. Shi, J. Phys. Chem. B \textbf{116}, 1164 (2012); P. K. Ghosh, A. Y. Smirnov, and F. Nori, J. Chem. Phys. \textbf{134}, 244103 (2011); P. K. Ghosh, A. Y. Smirnov, and F. Nori, J. Chem. Phys.\textbf{ 131}, 035102 (2009); G. Ritschel, J. Roden, W. T. Strunz, and A. Eisfeld, New J. Phys. 13, 113034 (2011).



\bibitem{Cao}J. S. Cao and  R. J.  Silbey, J. Phys. Chem. A 113, 13825 (2009); A. Kelly  and Y. M. Rhee,  J. Phys. Chem. Lett. 2, 808 (2011).

\bibitem{Christensson} N. Christensson, H. F. Kauffmann, T. Pullerits, T. Mancal,arXiv:1201.6325; A. W. Chin, J. Prior, R. Rosenbach, F. Caycedo-Soler, S. F. Huelga, M. B. Plenio, arXiv:1203.0776.
\bibitem{Gnutzman} For a review, see S. Gnutzman and U. Smilansky, Adv. Phys. \textbf{55}, 527
(2006).





\bibitem{matrix} J. Adolphs and T. Renger, Biophys. J. {\bf 91}, 2778 (2006).

\bibitem{Note} An additional (the eighth) pigment  was recently discovered in each subunit.
However, we ignore this pigment since the eighth pigment is only
loosely bound and it usually detaches from the others when the
system is isolated from its environment to perform experiments,
see Ref.\cite{FEBS}. In addition, the main conclusions obtained
here are essentially the same for both eight and seven pigments.

\bibitem{FEBS}A. Ben-Shem, F. Frolow, N. Nelson, FEBS Lett. \textbf{564}, 274 (2004); J. Wen, H. Zhang, M. L. Gross, and
R. E. Blankenship, Biochemistry
\textbf{50}, 3502 (2011); D. E. Tronrud, J. Wen, L. Gay, R. E. Blankenship, Photosynth. Res. \textbf{100} 79 (2009);
 M. Schmidt am Busch, F. Muh, M. El-Amine Madjet, and T. Renger, J. Phys. Chem.
Lett. \textbf{2},93 (2011).

\bibitem{Yi2012} X. X. Yi, X. Y. Zhang, and C. H. Oh, arXiv:1208.4671.


\end{thebibliography}
\end{document}